\documentstyle[preprint,tighten,floats,prd,aps]{revtex}
\input epsf

\newcommand{\be}{\begin{eqnarray}}
\newcommand{\ee}{\end{eqnarray}}

\begin{document}
\draft
\preprint{\vbox {\hbox{UAHEP986}
\hbox{IFP-764-UNC} 
\hbox{November 1998}
\hbox {hep-ph/9811326} } } 
\title{Neutrino Masses in R Parity Violating Supersymmetry}
\author{L. Clavelli\footnote{lclavell@bama.ua.edu} and P.H. Frampton
\footnote{frampton@physics.unc.edu}}
\address{$^*$Department of Physics and Astronomy, University of Alabama,\\
Tuscaloosa AL 35487}
\address{$\dagger$ Department of Physics and Astronomy, University of
North Carolina,\\
 Chapel Hill NC 27599-3255}
\maketitle
\begin{abstract}
We discuss a model for neutrino masses and mixings based on R
parity violating Yukawa couplings. The model requires no right-handed
neutrinos. It accommodates the SuperKamiokande data on
atmospheric neutrinos and successfully 
relates this data to the mass difference for solar neutrinos in 
the small-angle MSW solution. 
We obtain an unexpected testable pattern
for the three neutrino flavors.
\end{abstract}
\par
Recently the SuperKamiokande collaboration \cite{superK} presented
striking evidence for neutrino masses and mixings.
The data extend by more than six orders of magnitude the previously
known hierarchy among fermion masses. The result could
suggest the existence of right handed neutrinos as required, for
example, in models of leptons as a fourth color \cite{Pati} or in
left-right symmetric models
\cite{Mohap}.  The extreme smallness of the neutrino masses might
afford a window on the grand unification scale \cite{Ramond}
through the see-saw mechanism \cite{GRS,Yanagida}. The importance of these
questions makes it imperative to give wide latitude to the
discussion of alternative possibilities for understanding neutrino
masses. The literature is, as expected,
mushrooming\cite{N1,N2,N3,N4,N5,N6,N7,N8,N9,N10,N11,N12,N13,N14}.

{\it The Class of Models.}
In this paper we consider R parity violating (RPV)
supersymmetry-based models leading to calculable neutrino masses
We arrive at a specific pattern of masses and mixings
quite different from the see-saw mechanism.
Similar models have already been the subject of study by other authors
\cite{Borz,Chun,Kong,Mukhopadhyaya,bar1}. 

For
definiteness we work in the context of
gravity mediated supersymmetry (SUSY) breaking scheme (MSSM) with
additional RPV Yukawa couplings. Gauge mediated SUSY breaking is
also compatible with the mechanism we wish to discuss.
After this introduction we discuss a simple ansatz that
leads to the observed neutrino squared mass differences and the
apparent large mixing seen in the atmospheric experiments, and discuss
a slight generalization of the simplest model which retains the
same general features.
Then, we describe a Monte-Carlo search in the parameter
space of Yukawa couplings which favors this type of model 
and hence increases our confidence that our choices
are not arbitrary or {\it ad hoc}.
There follow the general constraints on the RPV couplings arising from
the neutrino observations, together with a summary and some 
pertinent general observations.

{\it The One-Loop Graph.}
For calculability of neutrino masses we assume the absence of right
handed neutrinos and tree level Majorana masses. The latter would be
forbidden, for example, in a model where there are no dimension three
operators in the SUSY breaking Lagrangian. Such a model would also
forbid gluino and photino masses at tree level and leaves open the
interesting possibility that the neutrinos could be mixing with the
photino. The minimum expectation for the photino mass, however, would
be in the MeV region far above the current range of apparent neutrino
masses so we do not consider this possible component in the mixing
matrix at the present time. Whether or not the gluino and photino are
light has no direct bearing on the current model. We, therefore, do
not discuss proposed counter-indications to the light gluino scenario
from LEP four jet analyses \cite{4jet} and counter-arguments thereto
\cite{Farrar}.  Similarly the LEP II indications from Chargino and
Higgs searches that the Higgs sector in the MSSM might have to be
enlarged by a singlet Higgs boson in the $m_{1/2}=A=0$ model is not
directly relevant to our present considerations.

We consider a Lepton number and R parity violating term in the
superpotential of the form
\be
   {\cal W} = \lambda_{ijk} e L^iL^j \bar {E}^k
\label{superpot}
\ee
where L represents the left handed lepton doublet superfield in the
i'th family and E represents the singlet lepton superfield of the k'th
family. The renormalization group behavior of this RPV coupling is
discussed in
\cite{Barger}.  Because of the superfield antisymmetry, the $\lambda 's$
are antisymmetric in the first two indices leaving nine independent
components. We have normalized them relative to the electric charge e.
The corresponding piece of the interaction Lagrangian that will
contribute to neutrino masses at one loop order is
\be
    {\cal L} = \lambda_{ijk} e \bigl( {\tilde e}^j_L \bar {e}^k_R
     \nu^i_L + {\tilde e}^{j*}_L (\bar {\nu}^i_L) e^k_L
     - i \leftrightarrow j \bigr )
\label{Lagrangian}
\ee
The loop graph of figure 1 yields the neutrino mass matrix
\be
    {\cal M}_{ij} = \sum_{k,l} \lambda_{ikl} \lambda^*_{jkl}
       { \alpha \over{ 3 \pi}} F(k,l) 
\label{Massmatrix}
\ee
\begin{figure}
\hskip 3.0cm
\epsfxsize=3.5in \epsfysize=2in \epsfbox{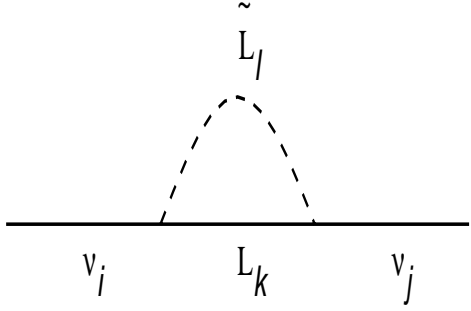}
\caption{R-parity violating graph leading to neutrino mixing}
\label{Fig.1}
\end{figure}
The function $F(k,l)$ is the same as appears in the one loop
contribution to gluino and photino masses \cite{BGM}. It can be
written in terms of the k'th family charged lepton and slepton masses
$m_k ,{\tilde m}_{k,R} ,{\tilde m}_{k,L}$ or in terms of the ratios
$x_{k,L} = \frac {m^2_k} {{\tilde m}^2_{k,L}} , x_{k,R} = \frac {m^2_k}
{{\tilde m}^2_{k,R}}$.
\be
    F(k,l) = m_l \bigl( (1-x_{k,R})^{-1} \ln (x_{k,R}) -
    (1-x_{k,L})^{-1} \ln (x_{k,L}) \bigr )
\label{F(k,l)}
\ee
We take the $\lambda_{ijk}$ to be real for the purposes of this paper,
ignoring possible CP violating phases. For the charged slepton masses
we use the supergravity related expectations
\begin{eqnarray}
   {\tilde m}^2_{k,L} & = &m_0^2 + m_k^2 + M_Z^2 \cos{2 \beta} \bigl[-1/2 +
       \sin^2{\theta} \bigr ] \\
   {\tilde m}^2_{k,R} & = & m_0^2 + m_k^2 - M_Z^2 \cos{2 \beta} \sin^2{\theta}
\label{sleptonmasses}
\ee
Then, neglecting higher order terms, $F(k,l)$ takes the
form
\be
    F(k,l) = \frac{M_Z^2}{M^2} \cos{2 \beta} (1/2 - \sin^2\theta )
              m_l \bigl(1 + \frac {m_k^2}{M^2} \ln\frac{m_k^2}{M^2}
               + {\cal O}(1/M^4) \bigr)
\label{Fapprox2}
\ee
where
\be
     M^2 = m_0^2 - \cos{(2 \beta)} M_Z^2/4
\label{Msq}
\ee
\par
To a good approximation the neutrino masses are proportional to the
the SUSY parameters in the combination $\lambda^2 \cos{(2 \beta)}
M_Z^2 /M^2$. We adopt the nominal values $\tan(\beta) = 1.5 , m_0 =
130 GeV$ but other possible values can be accommodated by rescaling
the $\lambda 's$ by the appropriate factor.

{\it The Simplest Model.}
The {\it simplest} model
discussed in this paper is that where 
$\lambda_{121}=\lambda_{131}=\lambda_0$ and all other linearly
independent $\lambda 's$ vanish. With this assumption which will be
further justified below one can fit the apparent neutrino
squared mass differences suggested by the atmospheric and solar
neutrino data and a large mixing in the $\nu_\mu - \nu_\tau$ channel
with moderate values of $\lambda_0$ and no right handed neutrinos or
GUT scale parameters. As discussed below, a discrete symmetry
yields this pattern of the $\lambda 's$. As we shall see,
whether or not the neutrino anomalies are due to these RPV couplings,
the current data provide strong experimental constraints on the values
of the independent $\lambda$'s. 

We first note that the main features of the neutrino data
can be fit in a simple one parameter model where this parameter is of
moderate size.  The model can be described by the as yet unmotivated
assumption that all the $\lambda$'s of Eq.(\ref{superpot}) vanish unless
two of the three indices lie in the first generation and that these
non-vanishing $\lambda 's$ have equal magnitude. That is we assume

\begin{equation}
\lambda_{121}=\lambda_{131}=\lambda_0;\quad \lambda_{ijk} = 0
\quad (otherwise)
\end{equation}

This pattern can be obtained by
imposing a $Z_2$ symmetry on the superpotential. Under this $Z_2$ ,
the right-handed superfields and the first generation left-handed
superfield changes sign. In order to provide for charged lepton
masses there must then be at least two Higgs doublets, one of which
couples to the first generation leptons and is even under the discrete
symmetry while the other(s), coupling to the second and third
generation leptons, are odd under the symmetry.

The non-zero elements
of the neutrino mass matrix are now:

\begin{equation}
M_{11} = \frac{\alpha \lambda_0^2}{3 \pi} (F(2,1)+F(3,1))
\end{equation}
and
\begin{equation}
M_{22} = M_{33} = M_{23} = M_{32} = \frac{\alpha \lambda_0^2}{3 \pi}
F(1,1)
\label{M's}
\end{equation}  
In this zeroth order model, it is apparent that one cannot describe
the mixing of the first generation (electron) neutrino. This mixing
must be attributed to other interactions or to residual effects from
other (small) $\lambda 's$. However, one can immediately see that this
mass matrix incorporates a $45 \deg$ mixing between the second and
third generation neutrinos and we can show that the squared mass
differences between the $\nu_\mu - \nu_\tau$ eigenstates and between
the $\nu_e$ and the heavier of the other two neutrinos are consistent
with observations for order unity values of $\lambda_0$.

Diagonalizing the above mass matrix, the three eigenvalues are
given by
\be
   m_1 = M_{11} \quad , \quad m_2 = 2 M_{33} \quad , \quad m_3 = 0
\label{eigvalues}
\ee
Substituting the expressions for $F(k,l)$ and neglecting
non-leading powers of the lepton masses relative to slepton masses,
the eigenstates have the squared mass differences

\begin{equation}
m_3^2 - m_2^2 = \left( \frac{\alpha \lambda_0^2}{3\pi} 
\left( \frac{m_Z^2}{M^2} \right) (1-4 \sin^2\theta) \cos 2\beta
\right)^2 m_e^2
\end{equation}

and

\begin{equation}
m_2^2 - m_1^2 = (m_3^2 - m_2^2) \left(
\frac{m_{\tau}^2}{M^2} \ln \frac{m_{\tau}^2}{M^2} \right)
\label{sqmdifs}
\end{equation}

For example the two squared mass splittings are given consistent with
experiment by reasonable SUSY parameters, $m_0 =130$ GeV,
$\cos(2\beta)=-.38$, if $\lambda_0$ is between $.1$ and $0.12$. If
$m_0$ ranges up to one TeV and tan$\beta$ decreases slightly toward
unity, the required $\lambda_0$ approaches unity. Independent of
$\lambda_0$ the model predicts a ratio of the two squared mass
differences in Eq.(\ref{sqmdifs})
consistent with the SuperKamiokande data and the MSW interpretation of
the solar data for the universal scalar mass $m_0$ in the $100$ GeV to
$1$ TeV range. The model disagrees with the vacuum mixing solution of
the solar neutrino anomaly since this would require a much smaller
value of the squared mass splitting ratio.

{\it An Extension of the Simplest Model.}
As an extension of the model, we obtain the same mass
matrix at leading order by allowing the more general
set of relationships:

\begin{eqnarray}
\lambda_{121} & = & \lambda_{131}\\
\lambda_{122} & = & \lambda_{132}\\
\lambda_{123} & = & \lambda_{133}\\
\lambda_{23k} & = & 0
\ee

It is straightforwardly checked that this maintains the results
of the previous subsection with the identification
\be
     \lambda_0 = \sum_{k} \lambda_{12k} m_k/m_e
\ee

{\it General Search of All Possible $\lambda_{ijk}$}
Here we discuss a general search in the parameter space
of the $\lambda_{ijk}$ for values consistent with the atmospheric and
solar neutrino experiments. Clearly, in order to fit the mass
splitting observations and the apparent large mixing angle between the
second and third eigenstates, the $\lambda 's$ must lie in the
neighborhood of the present simple model. However, in order to
obtain non zero mixing of the electron neutrino from the Yukawa
couplings, some small admixtures of the other $\lambda 's$ are required.
The allowed solutions are bounded by a nine dimensional rectangular
parallelopipoid
whose side lengths we find by a Monte-Carlo method.

We impose the antisymmetry requirement in the first two indices on the
$\lambda_{ijk}$ and, running over all possible values of the nine
independent components, we determine the neutrino mass matrix elements
from
Eq.(\ref{Massmatrix}).

Keeping the $\lambda_{ijk}$ real, the mass matrix is diagonalized by
an orthogonal matrix U satisfying
\be
         {\cal M} U = U {\cal M}_d
\label{U}
\ee
where ${\cal M}_d$ is the diagonal matrix of the three neutrino mass
eigenstates. With no right handed neutrinos and with a large mixing in
the $\nu_\mu-\nu\tau$ sector, there is no possibility of finding a large
mixing angle to $\nu_e$.  Thus we seek a solution corresponding to the
small angle MSW solution. A set of nine $\lambda_{ijk}$ is counted as a
solution if it satisfies the following experimental constraints
\parskip=0.pt \parindent=30pt
\begin{itemize}
\item(1)   $.0005 eV^2 < |m_2^2  - m_3^2 | < .006 eV^2$
\item(2)   $.5365 < |U_{22}| < .843  \quad (\sin(2 \theta)>< 0.83)$
\item(3)   $ |U_{21}| < 0.1 \sqrt(1-U_{22}^2)\quad$
(little mixing between $\nu_\mu$ and $\nu_e$)
\item(4)   $4\cdot 10^{-6} eV^2  < |m_1^2 - m_3^2 | < 1 \cdot 10^{-5}
eV^2$ (MSW solution)
\item(5)   $0.9985 < |U_{11}| < 0.99985$ (small angle MSW case)  
\end{itemize}
Without loss of generality we can choose one of the $\lambda's$ to be
positive.  We therefore choose $\lambda_{121} > 0$.  Similarly, reversing
the sign of all the $\lambda's$ with an odd number of index three's, has
the effect of interchanging the eigenstates in the $2-3$ sector.  We
therefore choose $\lambda_{131} > 0$ and label the heavier eigenstate in
the strongly mixed $2-3$ sector as $\nu_2$.

{\it Constraints on the RPV Interactions.}
Here we compare the above constraints 
with RPV constraints found in earlier studies\cite{bar1,bar2}.
The solutions to the five constraints listed above require
\begin{eqnarray}
0.0 & < \lambda_{121} & < 0.107\\
-5.7 \cdot 10^{-3} & < \lambda_{122} & < 6.7 \cdot 10^{-3}\\
-1.9 \cdot 10^-3 & < \lambda_{123} & < 1.6 \cdot 10^{-3}\\
-1.9 \cdot 10^{-2} & < \lambda_{231} & < 1.6 \cdot 10^{-2}\\
-8.8 \cdot 10^{-4} & < \lambda_{232} & < 1.2 \cdot 10^{-3}\\
-2.4 \cdot 10^{-4} &< \lambda_{233} & < 2.1 \cdot 10^{-4}\\
0.0 & < \lambda_{131} & < .106\\
-8.0 \cdot 10^{-3} & < \lambda_{312} & < 6.9 \cdot 10^{-3}\\
-2.0 \cdot 10^{-3} & < \lambda_{313} & < 1.7 \cdot 10^{-3}
\label{mcresults}
\ee
The upper limit on $\lambda_{121}$ obtained here is less than the
upper limit quoted earlier in the simple model due to the incorporation
here of the electron neutrino mixing angle constraints.

   If the neutrino mixing is primarily due to some mechanism other 
than that treated here,
our results can be taken as upper limits on the $\lambda's$ with
the results changing as discussed above if the other SUSY parameters
are varied.  The upper limits on the $\lambda's$ obtained here are
comparable to the upper limits obtained from other constraints such
as from charged current universality and from
$\tau$ decay \cite{bar2,CCLQ}. Thus the R-parity violating mechanism for
neutrino mixing will give interesting predictions for other areas
of phenomenology with only slight improvements in accuracy.
For instance the upper limit of \ref{mcresults} on $\lambda_{121}$
consistent with the neutrino data is close to the upper limit $0.17$
(in our normalization) obtained in \cite{bar2,CCLQ} from charged current
universality.

{\it Summary.}
Our main result is that a simple RPV pattern of couplings gives
rise to the $\nu_{\mu} - \nu_{\tau}$ mixing with near maximal
mixing consistent with the recent SuperKamiokande
data. Scaling that mass difference to the experimentally-
favored one of $\Delta m^2_{\mu\tau} \sim 0.0025 eV^2$
and putting in the mixing angle for $\nu_{e}-\nu_{\mu}$
of sin$^2 2\theta_{e\mu} \sim 0.01$ suggested by the small-angle
MSW effect, one is led to a squared mass difference of
$\Delta m_{e\mu}^2 \sim 10^{-5} eV^2$.  A distinctive prediction of
the current picture is that the mass of the electron neutrino is
close to the heavier of the other two neutrino masses whereas most
other models predict it closer to the lighter of the other two.

In fact the neutrino mass spectrum we predict is
{\it upside-down} with respect to the see-saw expectation.
Arbitrarily normalizing the mass eigenstate $\nu^{'}_{\mu}$
to $0.050000 eV$, our spectrum in the mass basis looks like:
$M(\nu^{'}_e) \simeq 0.050125eV$; 
$M(\nu^{'}_{\mu}) = 0.050000eV$; and 
$M(\nu^{'}_{\tau}) \simeq - 0.000125eV$. 
The mixing angles are approximately $\Theta_{\mu\tau} \simeq \pi/4$,
$\Theta_{e \mu} \simeq 0.05$ and, although it is not fully constrained
the third angle should be small $\Theta_{e \tau} 
\simeq \Theta_{\mu \tau} \simeq 0.05$. Of course, our subscripts
for $\nu^{'}_{\mu}$ and $\nu^{'}_{\tau}$ in the mass basis are only
suggestive, because both are approximately equal mixtures of
$\nu_{\mu}$ and $\nu_{\tau}$.

With this convention, the pattern which emerges for the squared neutrino mass differences is:
\begin{equation}
\Delta M^2_{\mu \tau} \simeq \Delta M^2_{\tau e} \gg \Delta M^2_{e \mu}
\label{Delta}
\end{equation}
while, at the same time, the neutrino mixing angles satisfy:
\begin{equation}
\Theta_{\mu \tau} \gg  
\Theta_{\tau e} \simeq  
\Theta_{e \mu}.
\label{Theta}
\end{equation}
It would be interesting if this 
prediction for the neutrino masses and mixings is borne out by further 
experiments. It is possible that Eq.(\ref{Delta}) and Eq.(\ref{Theta})
are more general than this model.

Finally we re-emphasize that we are assuming only the
minimal three left-handed neutrino fields $\nu_{iL}$ of the Standard Model.
Also, we note that the electron neutrino mass $M(\nu_e)$, while  
two orders of magnitude beyond current bounds from the tritium $\beta$- decay
end point, is only one order of magnitude beyond the
present empirical limit arising from $(\beta\beta)_{0 \nu}$, 
neutrinoless double $\beta$- decay.

This work was supported in part
by the US Department of Energy under Grant no. DE-FG02-96ER-40967 at
the University of Alabama and under Grant no. DE-FG05-85ER-40219
at the University of North Carolina.

%\begin{figure}
%\begin{center}
%\ \epsfbox{986fig1.eps}
%\caption[]{One-loop diagram.}
%\end{center}
%\end{figure}

\end{document}